\documentclass[preprint,authoryear,12pt]{elsarticle}%
\usepackage{graphicx}
\usepackage{epsfig}
\usepackage{amssymb}
\usepackage{amsthm}
\journal{New Astronomy}
\begin{document}
\begin{frontmatter}
\title{Population I Cepheids and star formation history of the Large Magellanic Cloud}
\author[ari]{Y. C. Joshi\corref{cor1}}
\ead{yogesh@aries.res.in}
\cortext[cor1]{Y. C. Joshi}
\author[ari]{S. Joshi}
\address[ari]{Aryabhatta Research Institute of Observational Sciences, Nainital, India 263 129}
\begin{abstract}
In this paper we study the Cepheids distribution in the Large Magellanic Cloud (LMC) as a function of their ages using data from the OGLE~III photometric catalogue. To determine age of the Pop I Cepheids, we derived a period - age (PA) relationship using the Cepheids found in the LMC star clusters. We find two peaks in the period distribution at $\log P =0.49\pm0.01$ and $\log P =0.28\pm0.01$ days which correspond to fundamental and first overtone pulsation modes, respectively. Ages of the Cepheids are used to understand star formation scenario in the LMC in last 30 -- 600\,Myr. The age distribution of the LMC Cepheids is found to have a peak at log(Age)=8.2$\pm$0.1. This suggests that major star formation event took place at about 125-200\,Myr ago which may have been triggered by a close encounter between the SMC and the LMC. Cepheids are found to be asymmetrically distributed throughout the LMC and many of them lie in clumpy structures along the bar. The frequency distribution of Cepheids suggests that most of the clumps are located to the eastern side of the LMC optical center.
\end{abstract}


\end{frontmatter}

\section{Introduction}
\label{sec:intro}
The Large Magellanic Cloud (LMC) is among one of the most studied galaxy in the Universe due to its close proximity to the Galaxy, favourable viewing angle, and star formation activities. In recent times, the distribution of stellar populations in the LMC has been studied with variety of objects, e.g. star clusters (Pietrzynski \& Udalski 2000, Harris \& Zaritsky 2009, Glat et al. 2010), Cepheid variables (Alcock et al. 1999, Nikoleav et al. 2004), RR Lyrae variables (Subramaniam \& Subramanian 2009, Wagner-Kaiser \& Sarajedini 2013), red clump stars (Koerwer 2009, Subramaniam \& Subramanian 2013), among others. These studies imply that the episodic star formation events have taken place in the LMC, most likely due to repeated interaction between the Magellanic Clouds (MCs) and/or with the Galaxy.

Population I Cepheids have been widely used to reconstruct the history of star formation because they are intrinsically bright, easily observable and ubiquitous in the LMC. They are very good candidates to understand the star formation activity during the last 30 -- 600\,Myrs as typical life of the Pop~I Cepheids lie in this time span. There are two principal types of Pop~I Cepheids, one those pulsating in the fundamental mode (FU) exhibit sawtooth-like light curves and other first overtone (FO) generally show sinusoidal light curves, have shorter periods and typically have lower amplitudes than the FU Cepheids. The use of Cepheids as tracers of young stellar populations comes from the fact that they obey a period-luminosity and a mass-luminosity relation.

There has been many surveys for Cepheids in the LMC, however, due to selection biases, most of them were not complete. In recent years the amount and quality of the photometric data has increased manifold, thanks mainly to microlensing surveys (e.g., Beaulieu et al. 1995, Alcock et al. 1997, Udalski et al. 1999). One such survey, the Optical Gravitational Lensing Experiment (OGLE), has revolutionized the field by producing thousands of Cepheid light curves with high signal-to-noise (S/N) and determining very accurate parameters like periods, magnitudes and amplitudes (Soszy\'{n}ski et al. 2008). In this paper we aim to understand the Cepheids period and age distributions in the LMC and to study the spatial distribution of the Pop~I Cepheids in order to examine the star formation scenarios in the LMC. The paper is organized as follows: details of the data are given in Section 2. The period and age distributions of Cepheids are studied in Sections 3 and 4 respectively. Their spatial distributions are given in Section 5. A comparative study of the spatial distribution of Cepheids with that of the star clusters is presented in Section 6 followed by the summary of the results in Section 7.

\section{Data}
\label{sec:data}

The OGLE-III survey acquired high quality photometric data in the $V$ and $I$-bands 
for stars in a 40 square degree region of the LMC using the 1.3-m Warsaw telescope 
at the Las Campanas Observatory, Chile. The imaging camera is composed of eight CCDs each with $2k\times4k$ pixels covering a field of view of $35\times35$ arcmin$^2$ of the sky. A detail of the reduction procedures, photometric calibrations, and astrometric transformations are given in the Udalski et al. (2008). A complete photometrc catalogue of their survey including the variable stars is available online. Recently, they have made available a catalogue of 3375 Classical Cepheids (CCs) containing 1849 FU, 1238 FO, 14 second-overtone and 274 multi-mode Cepheids. CCs were identified from their light curves and location in the period - luminosity diagram (Soszy\'{n}ski et al. 2008). The distinction between FU and FO Cepheids was made using their locations in the $W_I - \log P$ diagram (reddening-free Wesenheit index $W_I = I - 1.55(V - I)$), as well as using 
the Fourier parameters $R_{21}$ and $\phi_{21}$. The final catalogue provides 
coordinates of the Cepheids, their mean $V$ and $I$ magnitudes, amplitudes,
periods and classification as well as cross-identification with other
catalogues. In this work, we concentrate our study on a sample of 3087 CCs including 1849 FU and 1238 FO Cepheids as their ages can be determined accurately from their pulsation periods.

\section{Period distribution of the Cepheids}

\begin{figure}
\centering
\includegraphics[height=7.0cm,width=12.0cm]{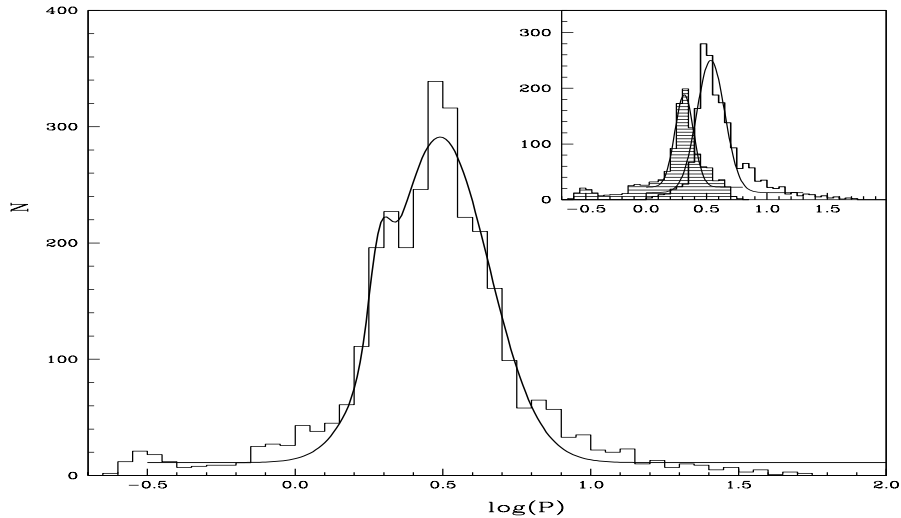}
\caption{The period distribution of Pop~I Cepheids as a function of pulsation 
period. A double Gaussian fit to the distribution is shown by a continuous line. 
In the top-right corner, we plot the individual period distributions for FU 
Cepheids and for FO Cepheids (shaded) and Gaussian profiles are shown
by continuous lines.} 
\label{figure:period_distri}
\end{figure}

In the OGLE-III catalogue, the periods, $P$ (in days), of FU Cepheids are in the range $0.01 < log P < 2.13$, while for FO Cepheids they are in the range $-0.60 < \log P < 0.77$. The error in the periods for more than 95\% of CCs is reported to be less than 0.001\% so we did not take it into account in the subsequent analysis. We found only 110 Cepheids which pulsate with $\log P > 1$, all of them in FU mode. It is well known that the number of Cepheids in any galaxy is not uniformly distributed over all the pulsation periods, and period distribution depends upon chemical composition along with many other factors like initial mass function, structure and evolutionary time scales of stars during their transit through the instability strip, etc (Alcock et al. 1999, Joshi et al. 2010). To understand the period distribution of CCs in the LMC, we plot the period distribution in Figure~\ref{figure:period_distri}, with a bin width of $\Delta\log P = 0.05$. It is quite evident that the period distribution of CCs shows a bimodal pattern. A combination of two Gaussian function fitted in the period distribution resulted in the two peaks at $\log P = 0.28\pm0.01$ and $0.49\pm0.01$, as shown by a continuous line in the same figure. The individual period distributions of FO and FU Cepheids are shown in the insect. We notice that these distributions are well represented by a Gaussian function and their peaks lie almost at the same positions as drawn in the bimodal distribution of the CCs. We therefore conclude that the two peaks seen in the period distribution actually corresponds to two frequency modes of Cepheids in which they pulsates.
%

Using the sample of 1360 CCs provided by the 
OGLE-I database, Antonello et al. (2002) did not report any bimodal distribution 
in the LMC. They found a peak at $\log P \sim 1.0$, however, on a closer inspection at their period distribution, we noticed a second peak at $\log P \sim 0.3$. It is 
worth mentioning here that Antonello et al. (2002) and Macri (2004) using
data from the DIRECT survey, and Joshi et al. (2010) using data from the 
Nainital Microlensing Survey, found two peaks at $\log P \sim 0.9$ and 1.1 
in the Cepheids period distribution in M\,31. For Milky Way Cepheids, 
Vilardell et al. (2007) and Joshi et al. (2010) found two peaks at $\log P= 0.7$
and 1.1. These studies suggest that the Cepheids period distribution shows a bimodal pattern in almost all the Local Group galaxies irrespective of the chemical composition of the host galaxy.

It is not clear whether the bimodal distribution found in the LMC as 
well as other nearby Local Group galaxies, such as the Milky Way, M\,31, M\,33 
and IC\,1613, is caused by (a) Cepheids pulsating in two different modes
(present study and Antonello et al. (2002)), (b) by non-detection of some 
Cepheids in a particular period interval due to lack of completeness 
(Joshi et al. 2010 and references therein), (c) pulsation at a particular 
period range caused by the instability of the non-linear fundamental pulsation 
cycle, particularly in metal rich galaxies (Buchler et al. 1997), or a combination of the above. However, 
the bimodal distribution seen in the LMC, which is a metal poor 
galaxy, does not support the instability interpretation of the Buchler et al. (1997). It is believed that evolution prior to and during the Cepheid pulsation phase, location in the
instability strip and metallicity are internal factors that determine the 
Cepheid period distribution. On the other hand, the star formation rate and 
initial mass function are external factors that dictates the frequency distribution of 
Cepheids as a function of period (Alcock et al. 1999). Some earlier studies
proposed that a two-component birth rate function may be responsible for the 
double peak in Cepheids period distribution (Becker, Iben \& Tuggle 1977),
however, Buchler et al. (1997) ruled out such possibility as the only condition to invoke
bimodal distribution.
Unless a homogeneous and near complete sample of Cepheids are obtained as well as identification of the pulsation modes, for different galaxies of the Local group, as has been done in the LMC by the OGLE 
survey, no satisfactorily conclusion can be drawn. 
Nevertheless, it is conspicuous from these studies (e.g, Antonello et al. 2002, Vilardell et al. 2007, Joshi et al. 2010) along with the present one that while metal poor galaxies (e.g, LMC) have peaks at smaller periods, metal rich galaxies (e.g, Milky way, M31) have peaks at longer periods and it is evident that maximum frequency in the period distribution shifted towards higher period with increase in the metallicity.

\section{Age distribution of the Cepheids}
\label{sec:ana}

Several empirical period - age (PA) relations for the Galactic, LMC, and M\,31 
Cepheids have been derived to estimate the age of the Cepheids. 
First such relationship for $\delta$-scuti stars was given by Tsvetkov (1986) which formulated semi-empirical PA and PAC relations. 
\begin{figure}
\centering
\includegraphics[height=7.0cm,width=10.0cm]{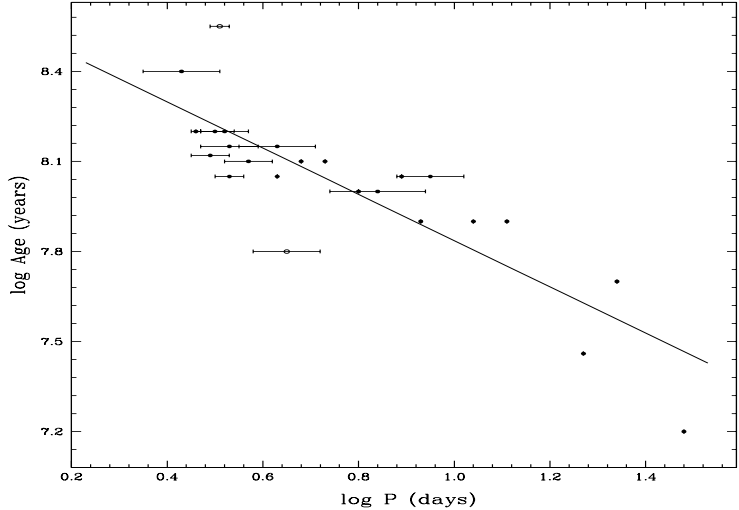}
\caption{The logarithms of the clusters ages and mean of the pulsation periods of the cluster Cepheids are plotted. A linear least square fit is drawn by a continuous line. Points shown by open circle are excluded from the fit.}
\label{figure:fig02}
\end{figure}
Later, Magnier et al. (1997) derived a PA relation using Cepheids in the NGC\,206 and its superposition in M\,31, to trace the age distribution and in turn, the star formation history in that region. However, the uncertainty in their $\log ({\rm Age})$ determination was more than 0.15. Subsequently, Efremov \& Elmegreen (1998) and Efremov (2003) have derived many PA relations considering different combinations of the 74 cluster Cepheids in 25 different open clusters in the LMC. They did not include Cepheids with small pulsation amplitudes and nearly sinusoidal light curves which could be either FO Cepheids or those at the first crossing of the the instability strip because their periods do not correspond to ages in the same way as for the FU Cepheids. However, their relationship has not been updated for the recent age determinations of the LMC clusters. We have therefore attempted to improve the PA relationship using the sample of Cepheids given by the Efremov (2003) but ages of the clusters taken from the recent study by Pandey et al. (2010) which has determined clusters ages from the integrated magnitudes and colours of the main-sequence population in the LMC. For each cluster, we determined a mean period of the Cepheids. We show the mean period against the cluster age in logarithmic scale in Figure~\ref{figure:fig02}. We draw a linear least square fit to find a relationship between log(Age) and logP as
\begin{figure}
\centering
\includegraphics[height=7.0cm,width=12.0cm]{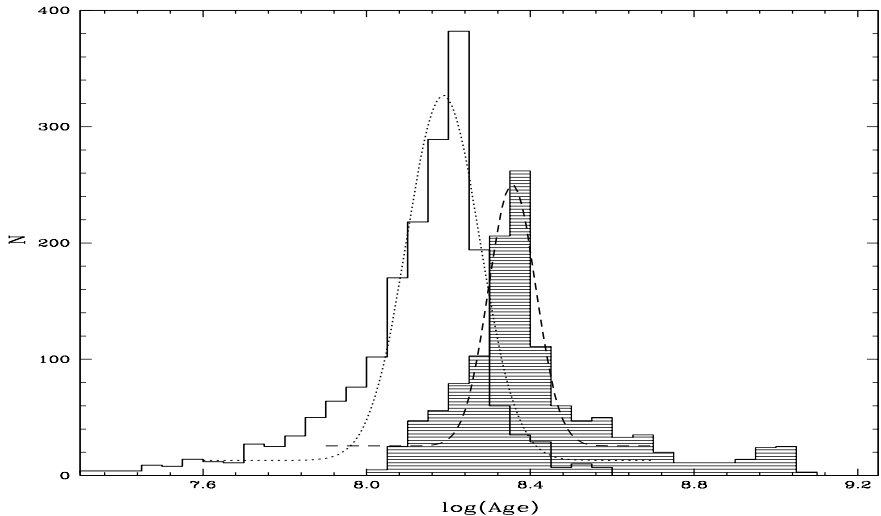}
\caption{Age distributions for the LMC Cepheids. The ages of the Cepheids were estimated using the PA relation derived in the present study. The continuous line represents FU Cepheids whereas dashed line in shades area represents FO Cepheids. The Gaussian profiles are shown by dotted and dashed lines for FU and FO Cepheids, respectively.}
\label{figure:fig03}
\end{figure}
\begin{eqnarray}
\log ({\rm Age}) = 8.60(\pm0.07) - 0.77(\pm 0.08)\log P
\end{eqnarray}
\noindent where Age is in years and P in days. The correlation coefficient of the fit is determined as 0.90. On the basis of evolutionary and pulsation models covering a broad range of stellar masses and chemical compositions, Bono et al. (2005) derived a PA relation for the metallicity of $z=0.01$ as $\log ({\rm Age}) = 8.41(\pm0.10) - 0.78(\pm 0.01)\log P$, with a standard deviation of 0.10. This shows a reasonable agreement is found between the PA relations derived in the present study for the LMC ($z=0.008$) with that of the Bono et al. (2005), considering the uncertainty in these relations as well as the difference in methods used to determine these relations. As pulsation period of Cepheids is the only quantity which can be determined from the observations of the pulsating stars with sufficiently high precision, their age can be determined with good approximation using the PA relation. Since the age is directly correlated with the pulsation period, hence Cepheids can be used to reconstruct the star formation history of the host galaxy within the life cycle of these intermediate-mass stars. For a given period we estimated age of each CC using the above relation. The resulting age distribution of CCs in a bin width of $\Delta(log(Age)) = 0.05$ is shown in Figure~\ref{figure:fig03}. Note that the age distribution of Cepheids show a pronounced peak followed by a sharp drop in frequency and FO Cepheids has a relatively longer tail after the maximum. The Gaussian fit in the age histogram of FU Cepheids shows a peak at $\log ({\rm Age}) = 8.19\pm0.01$, while FO Cepheids have a different distribution which is shifted towards higher ages and peaks at $\log ({\rm Age}) = 8.36\pm0.01$. Since these two populations show significantly different distributions, one cannot make any conclusion on the star formation in the LMC by combining them using a single PA relation.
%
\begin{figure}
\centering
\includegraphics[height=7.0cm,width=12.0cm]{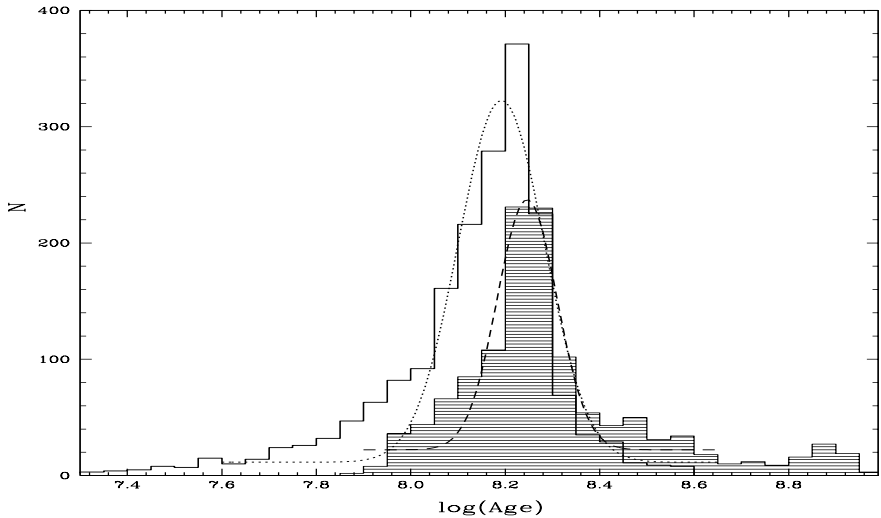}
\caption{The same as Figure~\ref{figure:fig02} but after converting 
the period of FO Ceheids to the corresponding period for FU Cepheids using 
Szil\'{a}di et al. (2007) relation.}
\label{figure:fig04}
\end{figure}

Period of FO Cepheids, however, can be transformed into the corresponding period for the FU Cepheids. Using the Cepheids detected in the LMC under the MACHO project, Alcock et al. (1995) found an empirical linear relation of the form ${P_1}/{P_0} = 0.733 - 0.034\log P_1$, where $P_0$ and $P_1$ are respective periods in FU and FO pulsation modes. Recently, Szil\'{a}di et al. (2007) took high-resolution spectroscopic observations of 17 double-mode Galactic Cepheids using 2.2-m telescope at La Silla observatory in Chile and derived a more robust linear relation
\begin{eqnarray}
\frac{P_1}{P_0} = 0.710(\pm0.001) - 0.014(\pm0.003) \log P_0 - 0.027(\pm0.004)[Fe/H]
\end{eqnarray}

We used above relation to transform the period of FO Cepheids to that of the equivalent period of FU Cepheids. Since Szil\'{a}di et al. (2007) found a weak dependence of [Fe/H] on the fundamental period as well as small range of metallicities, we ignored this correction in the transformation. Using the PA relation derived in the present study, we determined the Cepheids ages, and resulting age distribution is shown in Figure~\ref{figure:fig04}. It is quite clear that the age distributions for FU and FO Cepheids after period conversion resembles with each other and show peaks at $8.19\pm0.01$ and $8.25\pm0.01$ for FU and FO Cepheids, respectively, where errors are the statistical error in the peak age estimation in the Gaussian fit. Considering the error in log(Age) determination in equation (1), these two values are well in agreement.

\begin{figure}
\centering
\includegraphics[height=7.0cm,width=12.0cm]{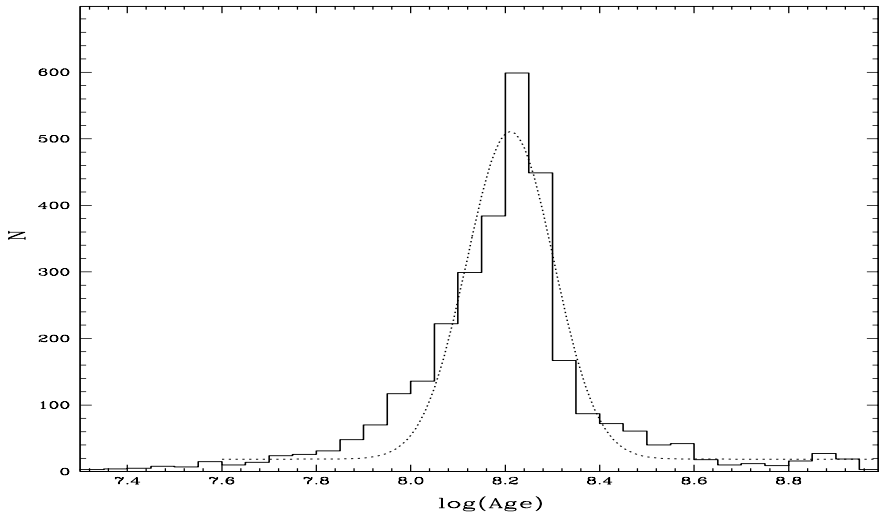}
\caption{Age histogram of the combined data using the PA relation after converting period of FO Cepheids to the corresponding period for FU Cepheids. The dotted line shows the best fit Gaussian profile.}
\label{figure:fig05}
\end{figure}

When both types of Cepheids are combined in Figure~\ref{figure:fig05}, the peak in the Gaussian fit was found at $\log({\rm Age}) \sim 8.2\pm0.1$, which is consistent with the age distribution of both FU and FO Cepheids individually. Here, quoted uncertainty in the peak age represents a combined error resulted due to the statistical error in peak age estimation in the Gaussian fit and the errors on the coefficients of equation (1) in the PA relation. The enhancement of CCs around 125-200\,Myrs ago hints of a major star formation episode at that time, possibly due to a close encounter between two segments of the Magellanic Cloud. The models proposed by Bekki \& Chiba (2005) and Kallivayalil et al. (2006) show that the last close encounter between the SMC and the LMC occurred about 100-200\,Myrs ago that has triggered star formation activity both in the LMC and the SMC. Using the open star clusters, Pietrzynski \& Udalski (2000) has studied the star formation history in the Magellanic clouds and detected three periods of enhanced cluster formation at about 800\,Myr, 125\,Myr, and 7\,Myr. A similar study by Glatt et al. (2010) based on the star clusters has found two peaks of enhanced cluster formation at the ages of 800\,Myr and 125\,Myr. The peak in the age distribution between the present study with those determined through star clusters in previous studies are thus in good agreement within the given uncertainties. The enhanced freqency of stars and star clusters in the Magellanic Clouds are widely believed to be due to the encounters between the SMC and the LMC. The increase or decrease in star formation rates depend upon whether these two Magellanic Clouds are approaching or receding. A repeated tidal interaction between these two clouds thus lead to the episodic star formation events in both the dwarf galaxies. It is, however, pointed out in some earlier studies (e.g., Glatt et al. 2010, and references therein) that not only the cloud-cloud interaction but stellar winds and supernova explosions can also induce such star formation events in the Magellanic Clouds.
\section{Spatial distribution of Population I Cepheids}
%
\begin{figure}
\centering
\vspace{-1.0cm}
\includegraphics[height=8.5cm,width=14.0cm,angle=0]{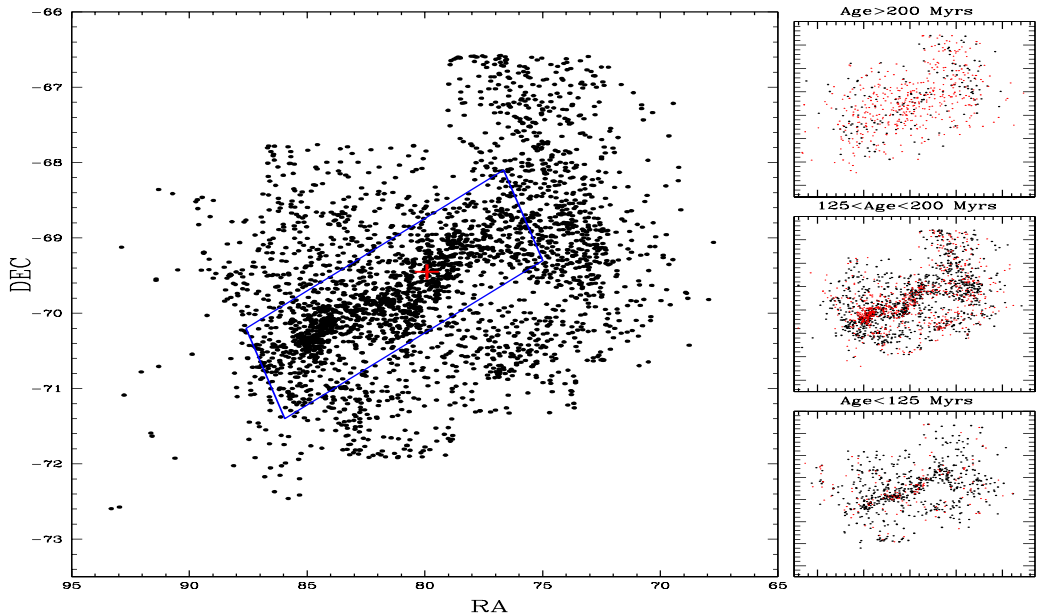}
\vspace{-1.0cm}
\caption{Spatial distribution of the CCs in the LMC from the OGLE III survey. The bar region is shown by a rectangular area. The plus sign in red colour represents the optical center of the LMC ($\alpha = 5^h19^m38^s$ and $\delta = -69^0 27^{'} 5^{''}.2)$ found by de Vaucouleurs \& Freeman (1972). In the right panels, the distribution of FU and FO Cepheids for 3 different age intervals are marked with black filled and red triangles, respectively.}
\label{figure:fig06}
\end{figure}
When we plot the distribution of Cepheids in the RA-DEC plane in Figure~\ref{figure:fig06}, it is apparent that the Cepheid distribution is not uniform across the LMC. Apart from the bar structure, the area surrounding the bar is also densely populated. The bar region and the optical center ($\alpha = 5^h19^m38^s$ and $\delta = -69^0 27^{'} 5^{''}.2$) given by de Vaucouleurs \& Freeman (1972) is also shown in Figure~\ref{figure:fig06}. It is noticed that optical center does not exactly coincide with the center of the Cepheids distribution but shifted away towards northwest direction. In the right panels of Figure~\ref{figure:fig06}, we draw similar distributions for three different age intervals that are $<$125\,Myr, 125-200\,Myr, and $>$200\,Myr. These values represent less than, within, and more than 1$\sigma$ value, respectively, of the maximum age determined from the age distributions of CCs in the previous section. We found that majority of the CCs have ages in the small range of 125-200\,Myr ($\sim$60\%). From the individual distribution of FU and FO Cepheids in the right panels, we see that the majority of FU Cepheids lie in the lower age group, whereas, FO Cepheids are dominant towards higher ages.

\begin{figure}
\centering
\vspace{-2.0cm}
\includegraphics[height=10.5cm,width=15.5cm,angle=0]{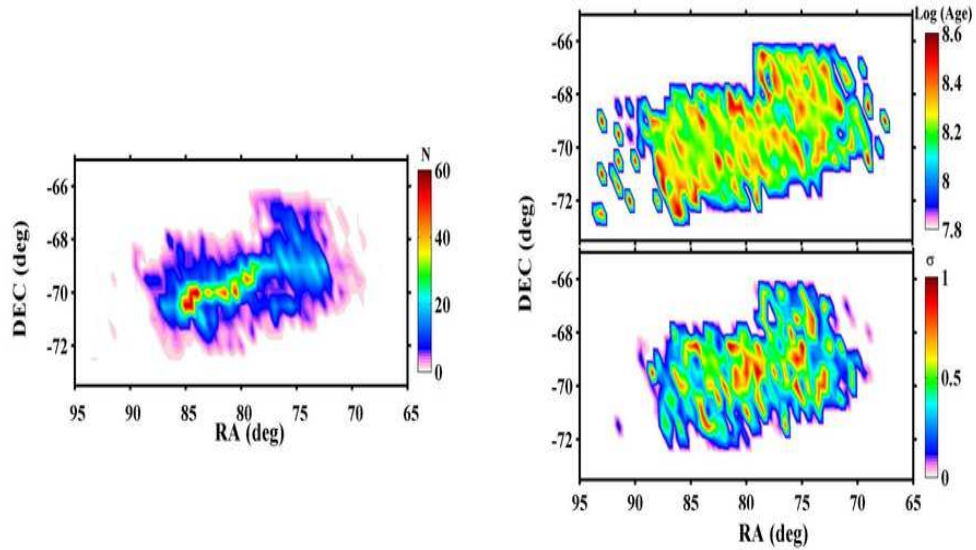}
\vspace{-2.5cm}
\centering
\caption{The map of CC distribution in the RA-DEC plane with two different bin sizes of $0.5 \times 0.5$ deg$^2$. The left panel shows frequency distribution, top-right panel shows the $\log({\rm Age})$ distribution, and bottom-right panel shows map for the dispersion in $\log({\rm Age})$.}
\label{figure:fig07}
\end{figure}

In order to study the star formation scenario in the radial direction of the LMC, we binned the CCs in the box sizes of $0.5 \times 0.5$ deg$^2$ which resulted in 365 boxes containing at least one CC. For each box, we determined the total number, mean age and dispersion in ages of the CCs lying within it. The map of the CCs distributions in RA-DEC plane is shown in Figure~\ref{figure:fig07}. In the left panel, we show the frequency distribution map for the CCs. It is seen that the bar region has the highest number of CCs and it is elongated in the east-west direction. The CCs are non-uniformly distributed along the bar and there are some pockets where clumpy structures are quite visible in the maps. The clumpy structure is more prominent in the southeast region, far-off from the optical center of the LMC. The freqency population of CCs, is minimum at the western part of the LMC and, in general, increases from northwest ($\alpha \approx 78^0, \delta \approx -69^0$) to the southeast ($\alpha \approx 85^0, \delta \approx -71^0$) region along the bar.
%

The mean $\log({\rm Age})$ distribution of CCs within each box in the range $7.8 < \log({\rm Age}) < 8.6$ is shown in the top-right panel of Figure~\ref{figure:fig07}. It shows that age distribution of the CCs is not same in all the regions within the LMC and there are plenty of small size structures where mean age is relatively higher, as represented by the red colour regions. Recently, Indu \& Subramaniam (2011) has studied last star formation event (LSFE) in the LMC by determining the Main-Sequence (MS) turn-off point in the colour-magnitude diagram which was found to be varied between 0-120\,Myr in various subregions. In their analysis, they found inner regions near the LMC bar have ages in the range of 0-40\,Myrs whereas peripheral regions have ages in the range of 60-100\,Myr. It is interesting to note that there is neither a gradual change in the age of CCs from inner region to outer region region as has been noticed in the LSFE map by Indu \& Subramaniam (2011), nor any preferential distribution of Cepheids in the eastern and western region of the galaxy. However, small scale asymmetry is still visible in some pockets within the LMC. This is further confirmed by the maps shown in bottom-right panel of Figure~\ref{figure:fig07} where dispersion in age of Cepheids within the boxes are shown. In general, we found small dispersion in most of the regions except few regions where dispersion is relatively larger. The higher dispersion in age does not correlate with the total number of CCs that suggests substantial star formation occurred over a short period of time at smaller scales and star formation proceeds in small regions faster than in large regions. If that was not the case, the dispersion in age would be larger where clumpy structure was found.
%
\begin{figure}
\vspace{-0.5cm}
\includegraphics[height=8cm,width=13.0cm,angle=0]{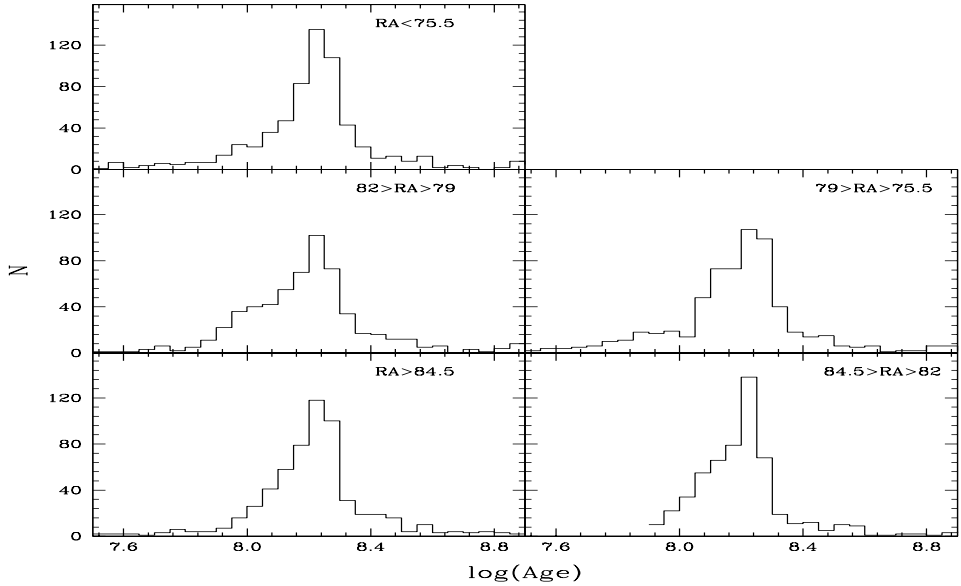}
\caption{Age distribution of the Cepheids in five different subregions in the LMC marked at the top of each panel.}
\label{figure:fig08}
\end{figure}

To study the age distribution of CCs across the LMC bar, we divided the sample of Cepheids into five different regions from East to West in the bar. The different sizes for the subregions are chosen keeping in mind that almost equal number of CCs fall in each region as well as number of CCs are statistically significant in order to draw any conclusion from the age distributions. Figure~\ref{figure:fig08} shows the age distributions of Pop~I Cepheids in the five subregions. It is evident from the plot that while the shape of the age distributions is different, they drop sharply at the same age in all the subregions. This peak age of CCs is closely associated with the metal abundance in the region. It is normally found that the peak age in the age distribution moves towards larger value with the increase of the metallicity. The distributions of CCs in various subregions across the LMC bar, as shown in Figure~\ref{figure:fig08}, suggest that the mean abundance does not significantly change across the bar. This is in agreement with the conclusion made by Alcock et al. (1999) after analysing period distributions of about 1800 Cepheids found in the LMC under the MACHO microlensing survey. Olsen \& Salyk (2002) and Grocholski et al. (2006) also did not find any metallcity gradient in the central region across the LMC from their analysis of red clump stars and star clusters, respectively.
\section{Comparison with the spatial distribution of star clusters}
\begin{figure}
\centering
\vspace{-10.9cm}
\includegraphics[height=28cm,width=17.0cm,angle=0]{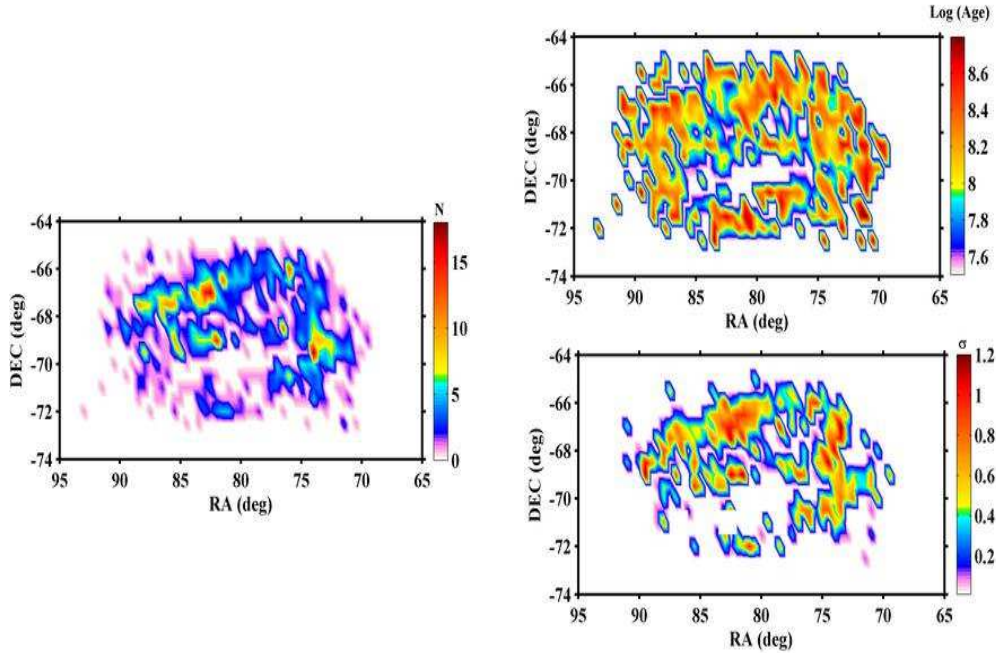}
\vspace{-10.8cm}
\caption{Same as Figure~\ref{figure:fig07} but for the star clusters taken from the Glatt et al. (2010).}
\label{figure:fig09}
\end{figure}

The spatial distributions of different populations in the LMC are well studied in recent times (Glatt et al. 2010, Wagner-Kaiser \& Sarajedini 2013, Subramaniam \& Subramanian 2013) and it is interesting to compare the spatial distribution between two different class of objects in order to understand the dominant mode of the large scale star formation within the galaxy at different timescales. One of the ideal candidates to compare with the spatial distribution of Cepheids in the LMC is star clusters as most of these populations belong to an age interval of log(Age) $\sim$ 7.5-8.5, a similar time scale in which most of the Cepheids are found to exist. This is more relevant in our case as age of the Cepheids in the present study has been determined using the star clusters itself. Glatt et al. (2010) recently published a comprehensive catalogue of star clusters providing accurate positions and ages for the 1194 clusters in the LMC. In Figure~\ref{figure:fig09}, we draw a similar spatial distribution of clusters as has already been drawn for the CCs. Here, we also made a box size of $0^{'}.5 \times 0^{'}.5$. The frequency distribution map shows that the star clusters are not homogeneously distributed throughout the LMC. They are either formed in the northeast or southwest regions and seem to avoid the bar region of the LMC. Studying the LSFE in the LMC, Indu \& Subramaniam (2011) also found that the recent star formations are stretched in the northeast direction. 

If we compare frequency distribution map of the cluster with that of the Cepheids, we see that the clumps of CCs do not coincide with the clumps of the star clusters. This is more visible when we compare them in the age interval of 125-200\,Myr where most of CCs were found in our data. A mutual avoiding of clumps of the Cepheids and star clusters in the LMC has also been noticed by Battinelli \& Efremov (1999). A possible reason for the different spatial distribution of star clusters and isolated stars such as Cepheids could be related to different modes of star formation with which they have been formed. This may answer why despite having thousands of clusters known in the LMC, only very few clusters are known to have Cepheids. The age map shows that old clusters belong to the outer region whereas younger clusters lie towards inner region of the LMC. This suggest an inwards quenching of star formation in the clusters. A similar scenario has been noticed by Indu \& Subramaniam (2011) from the study of LSFE using the data taken from both the OGLE III and Magellanic Cloud Photometric Survey (MCPS). We found maximum age of the age distribution of clusters around 100-125\,Myr which is in agreement with the Glat et al. (2010) who reported a similar peak at 125\,Myr. Taken together, we can conclude that there was indeed a star formation in the LMC took place at around 125\,Myr. In the age distribution determined through LSFE in the age interval of 0-120\,Myr, Indu \& Subramaniam (2011) noticed a hint of star formation at 90-100\,Myr. However, ages determined through the LSFE is mainly distributed in the smaller ages (0-80\,Myr) and only a small bump is seen in the age interval 90-100\,Myr, therefore, one can not make a conclusive remark on the basis of LSFE data alone. From the age dispersion map in the figure, dispersion is found to be correlated with the size of the clumps in the sense that larger clumps have higher dispersion in the clusters ages. Such a correlation was not found in the case of CCs.
\section{Discussion and Conclusions}\label{sec:discuss}

It is believed that close encounters between the Magellanic Clouds and the Galaxy, and/or between the two segments of the Magellanic Clouds has induced episodic star formation in these galaxies. Star formation in the LMC occurs on a large scale, both in terms of total number of stars formed and in the spatial extent. Pietrzynski \& Udalski (2000) found three periods of enhanced cluster formation in the LMC at about 7\,Myr, 125\,Myr, and 800\,Myr, in agreement with an earlier finding of Girardi et al (1995) from an age calibration of integrated colours of star clusters. From a detailed analysis of 20~million stars in the MCPS survey, Harris \& Zaritsky (2009) showed that enhanced star formation activities took place in the LMC at roughly 2 Gyr, 500\,Myr, 100\,Myr, and 12\,Myr. Glatt et al. (2010) studied 1193 star clusters in the LMC and found enhanced star formation at 125\,Myr and 800\,Myr. All these studies are broadly in agreement. Since it is well known that the pulsation period of a Cepheid is a good indicator of its age, Cepheids can therefore be used to reconstruct the star formation history in the LMC. However, as the maximum age of Pop~I Cepheids is not expected to be more than 500 -- 600\,Myr, our study of star formation using these variables provides no information about the details of the star formation episodes in the LMC outside this limited time period. 

The OGLE catalogue has allowed us to statistically analyse the star formation scenario in the LMC from very accurate period determinations of 1849 FU and 1238 FO Cepheids detected in their third phase of observations. Taking advantage of a large and homogeneous sample of CCs, we have studied the period distribution of CCs and found a bimodal period distribution. Our study shows that these two peaks actually correspond to FU and FO mode Cepheids showing maxima in the distribution at $\log P = 0.49\pm0.01$ and $0.28\pm0.01$, respectively. When we combined these two class of pulsating stars, after converting periods of FO Cephieds to that of the corresponding FU Cepheids, and employing a period-age relation derived through the LMC cluster Cepehids, we found an age distribution comprising of only a single age maxima with a pronounced peak at the log(Age)=8.2$\pm$0.1. Our study shows that the Cepheids are not homogeneously distributed throughout the LMC bar but lie in the clumpy structures. The clumps are more prominent in the southeast region, far-off from the optical center of the LMC. The enhanced population of Cepheids suggests that the a major star formation episode has taken place at around 125-200\,Myr ago in the LMC. However, a combined analysis of CCs and star clusters put this value at around 125\,Myr. It is believed that star formation episode at this time scale was triggered due to close encounter between the SMC and the LMC. On a comparison of spatial distribution of CCs and star clusters, a mutual avoiding of clumps of the CCs and star clusters was noticed in the LMC.

\section*{Acknowledgments}

We are thankful to the referee for constructive comments which has improved the paper. We are grateful to Dr. D. V. Phani Kumar for his assistance that helps to improve the presentation of the results. Thanks to Drs. Luis Balona and Chris Engelbrecht for their valuable inputs. This publication makes use of the catalog by the OGLE microlensing survey.

\end{document}